\documentclass[reprint,aps,prl,floattix]{revtex4-1}

\usepackage{amsmath}
\usepackage{amssymb}
\usepackage{amsfonts}
\usepackage{graphicx}
\usepackage{float}
\usepackage{psfrag}

\begin{document}

\title{Ultrafast X-ray spectroscopy of conical intersections}

\author{Simon P. Neville$^{1,2}$}
\author{Majed Chergui$^2$}
\author{Albert Stolow$^{1,3,4}$}
\author{Michael S. Schuurman$^{1,3}$}
\affiliation{$^1$Department of Chemistry and Biomolecular Sciences,
  University of Ottawa, 10 Marie Curie, Ottawa, Ontario, K1N 6N5,
  Canada}
\affiliation{$^2$Ecole Polytechnique F\'{e}d\'{e}drale de Lausanne,
  Laboratoire de Spectroscopie Ultrarapide and Lausanne Centre for
  Ultrafast Science (LACUS), Facult\'{e} des Sciences de Base, ISIC,
  Lausanne CH-1015, Switzerland}
\affiliation{$^3$National Research Council of Canada, 100 Sussex
  Drive, Ottawa, Ontario K1A 0R6, Canada}
\affiliation{$^4$Department of Physics, University of Ottawa, 150
  Louis Pasteur,Ottawa, ON K1N 6N5 Canada}

\begin{abstract}
  Ongoing developments in ultrafast X-ray sources offer powerful new
  means of probing the complex non-adiabatically coupled structural
  and electronic dynamics of photoexcited molecules. These
  non-Born-Oppenheimer effects are governed by general electronic
  degeneracies termed conical intersections which play a key role,
  analogous to that of a transition state, in the electronic-nuclear 
  dynamics of excited molecules. Using high level ab initio quantum dynamics simulations,
  we studied time-resolved X-ray absorption and photoelectron
  spectroscopy (TRXAS and TRXPS, respectively) of the prototypical
  unsaturated organic chromophore, ethylene, following excitation to
  its $S_{2}(\pi\pi^{*})$ state. The TRXAS in particular is highly sensitive to all aspects
  of the ensuing dynamics. These X-ray spectroscopies provide a clear
  signature of the wavepacket dynamics near conical intersections,
  related to charge localization effects driven by the
  nuclear dynamics. Given the ubiquity of charge localization in
  excited state dynamics, we believe that ultrafast X-ray
  spectroscopies offer a unique and powerful route to the direct
  observation of dynamics around conical intersections.
\end{abstract}

\maketitle

In excited states of polyatomic molecules, conical intersections (CIs)
play a central role, analogous to transition states in ground state
dynamics. The spatial range of the conical intersection dynamics is 
somewhat extended beyond the CI (a point of electronic degeneracy) itself. 
In the following, we use the general term conical intersection to
refer to this spatial region. Despite their widespread use in rationalizing ultrafast
electronic-nuclear dynamics\cite{ci-book}, unambiguous experimental
observation of CIs remains elusive. Here we investigate the use of
ultrafast X-ray spectroscopies to directly observe dynamics at
CIs. This is motivated by recent technological advances which result
in the availability of powerful ultrafast light sources in the X-ray
regime. The use of X-rays is appealing because they offer an
atom-specific probe of electronic and structural dynamics. While most
X-ray spectroscopic studies of molecular dynamics were initially
performed in the hard X-ray
range\cite{chergui_science_femtosecond_xanes,Lemke2013,Zhang2014},
studies have also recently appeared in the soft X-ray
range\cite{Huse2011}, in particular using table-top
sources\cite{Attar2017,Worner2017}.

There is a growing number of theoretical proposals for the use of
X-ray spectroscopies to probe the ultrafast nonadiabatic dynamics in
photoexcited molecules\cite{Capano2015,
  ethylene_faraday_ours,adc_xas_ours, Mukamel2015,
  mukamel_furan_x-ray_raman,
  mukamel_multidimensional_spectroscopy_cis}. Of these, only two techniques, 
namely Time-Resolved X-ray Absorption Spectroscopy (TRXAS) and 
Time-Resolved X-ray Photoelectron Spectroscopy (TRXPS), have 
demonstrated experimental feasibility and are therefore 
the focus of the present study. We recently showed that
the pre-edge region of the X-ray absorption spectrum offers a uniquely
sensitive probe of dynamics in the most fundamental unsaturated
hydrocarbon, the planar C$_2$H$_4$ molecule
ethylene\cite{ethylene_faraday_ours}. Specifically, the computed
time-resolved Carbon K-edge absorption spectrum contained
clear signatures of the excited state wavepacket
dynamics\cite{adc_xas_ours,ethylene_faraday_ours}.  Here we apply this
approach to the spectroscopy of conical intersections, an important
goal of ultrafast molecular sciences. In our previous work
on valence shell time-resolved photoelectron spectroscopy (TRPES), we
showed that the outgoing photoelectron is a particularly insightful
probe of nonadiabatic dynamics in
molecules\cite{stolow_trpes_chem_rev,Schalk_2011,allene2016}.
Therefore, we also simulate here the inner shell time-resolved X-ray
photoelectron spectrum (TRXPS) for the same photo-initiated
process. We believe that this comparison will help design the next
generation of X-ray spectroscopy experiments which are just
emerging\cite{Attar2017,Worner2017,Marangos2016}.

Via our simulations, we have uncovered a key advantage of 
ultrafast X-ray spectroscopy: that charge
localization effects in photo-excited ethylene lead to large (few eV)
splittings in the excited state pre-edge X-ray absorption spectrum.
This, in turn, offers a unique and acute sensitivity to wavepacket
dynamics at a CI. This observation has far reaching consequences,
given the commonality of charge localization in excited state
molecular dynamics and biological processes such as proton and electron transfer.

Our methods for calculating TRXAS were described in detail
elsewhere\cite{ethylene_faraday_ours}. Likewise, our simulations of
TRXPS are directly analogous to previously described methodologies for
computing valence shell electron (UV) TRPES\cite{Hudock2007}. 
Accordingly, we give only a brief summary of
the computational methodology here, with further details offered in 
the Supplemental Information. The {\it ab initio} multiple
spawning (AIMS) method\cite{martinez_fms_advchemphys} was used to
describe the time evolution of the excited state wavepacket. In the
AIMS method, the molecular wavefunction is expanded in a set of
adiabatic electronic functions,
$\left\{ | I(\boldsymbol{r}; \boldsymbol{R}) \rangle \right\}$, and
frozen Gaussian nuclear basis functions,
$\left\{ | g_{j}^{(I)}(\boldsymbol{R},t) \rangle \right\}$, of the
Heller form:

\begin{equation}
  | \Psi(\boldsymbol{R}, \boldsymbol{r}, t) \rangle = \sum_{I=1}^{n_{s}}
  \sum_{j=1}^{N_{I}} C_{j}^{(I)}(t) | g_{j}^{(I)}(\boldsymbol{R},t) \rangle | I
  (\boldsymbol{r}; \boldsymbol{R}) \rangle.
\end{equation}

The positions and momenta of the Gaussian basis functions evolve
according to the classical equations of motion, and the expansion
coefficients are evolved variationally via the solution of the
time-dependent nuclear Schr\"{o}dinger equation. The AIMS equations of
motion were solved on-the-fly using the results of electronic
structure calculations performed at the multireference first order
configuration interaction (MR-FOCI) level of theory. The details of
these calculations are given in the Supplementary Information.

In Figure \ref{fig:pes} we schematically shows the relevant electronic
states as a function of the torsional angle between the two planes
defined by each of the CH$_2$ groups in the molecule. Following
vertical excitation of the $S_{2}(\pi\pi^{*})$ state, the wavepacket
is localized at the structural and electronic character labeled
``A''. In line with previous studies\cite{mori_ethylene,
  champenois_ethylene_trpes, suzuki_ethylene, allison_ethylene_II}, a
rapid evolution of the initially excited $S_{2}$ state occurs by large
amplitude twisting motion about the C-C bond, leading to a change in
electronic character. The $S_{1}$ state is transiently populated and
has an electronic structure characterized by both valence
$\pi \pi^{*}$ and Rydberg $\pi 3s$ character (point ``B''). The total
excited state lifetime was determined to be $\sim$95 fs.  Internal
conversion to the $S_{0}$ ground state is found to occur via so-called
twisted-pyramidalized (Tw-Py) CI seam (point ``CI''). This CI
structure involves a 90$^{\circ}$ twist about the C-C double bond
combined with pyramidalization at one of the C atoms.

\begin{figure}
  \begin{center}
    \includegraphics[width=6.0cm,angle=0]{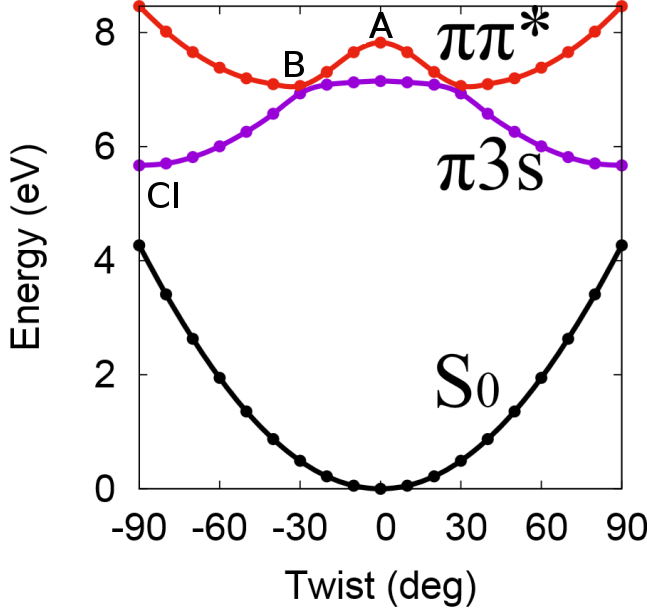}
    \caption{Relevant electronic states in ethylene as a function of
      the torsional angle about the C-C bond. The dynamics simulation
      initializes the wave packet at the FC geometry on the
      $S_{2}(\pi \pi^{*})$ state (point ``A''). The crossing between
      the initial $\pi\pi^{*}$ state and the $\pi3s$ state is denoted
      ``B''. The important conical intersection with the ground state
      is labeled ``CI'' and is the dynamical gateway (``transition
      state'') for the excited state.}
    \label{fig:pes}
  \end{center}
\end{figure}

We first examine the ability of TRXAS to probe this coupled electronic
and nuclear dynamics. The pre-edge part of the TRXAS, $\sigma(E,t)$,
was calculated as an incoherent sum over the transient spectra at the
centres of the Gaussian basis functions:

\begin{equation}
  \sigma(E,t) = \sum_{I=1}^{n_{s}} \sum_{j=1}^{N_{I}} \left|
    C_{j}^{(I)}(t) \right|^{2} \sigma_{I}\left( E;
    \bar{\boldsymbol{R}}_{j}^{(I)}(t) \right).
\end{equation}

\noindent
Here, $ \sigma_{I}\left( E; \boldsymbol{R}_{j}^{(I)}(t) \right)$
denotes the X-ray absorption spectrum (XAS) for the $I$th electronic
state calculated at the centre $ \bar{\boldsymbol{R}}_{j}^{(I)}(t)$ of
the Gaussian basis function $g_{j}^{(I)}(t)$. The XAS were calculated
using the second-order algebraic diagrammatic construction method
within the core-valence separation
approximation\cite{dreuw_cvs-adc3_2015} (CVS-ADC(2)) and the
6-311++G** basis.

\begin{figure}
  \begin{center}
    \includegraphics[width=7.0cm,angle=0]{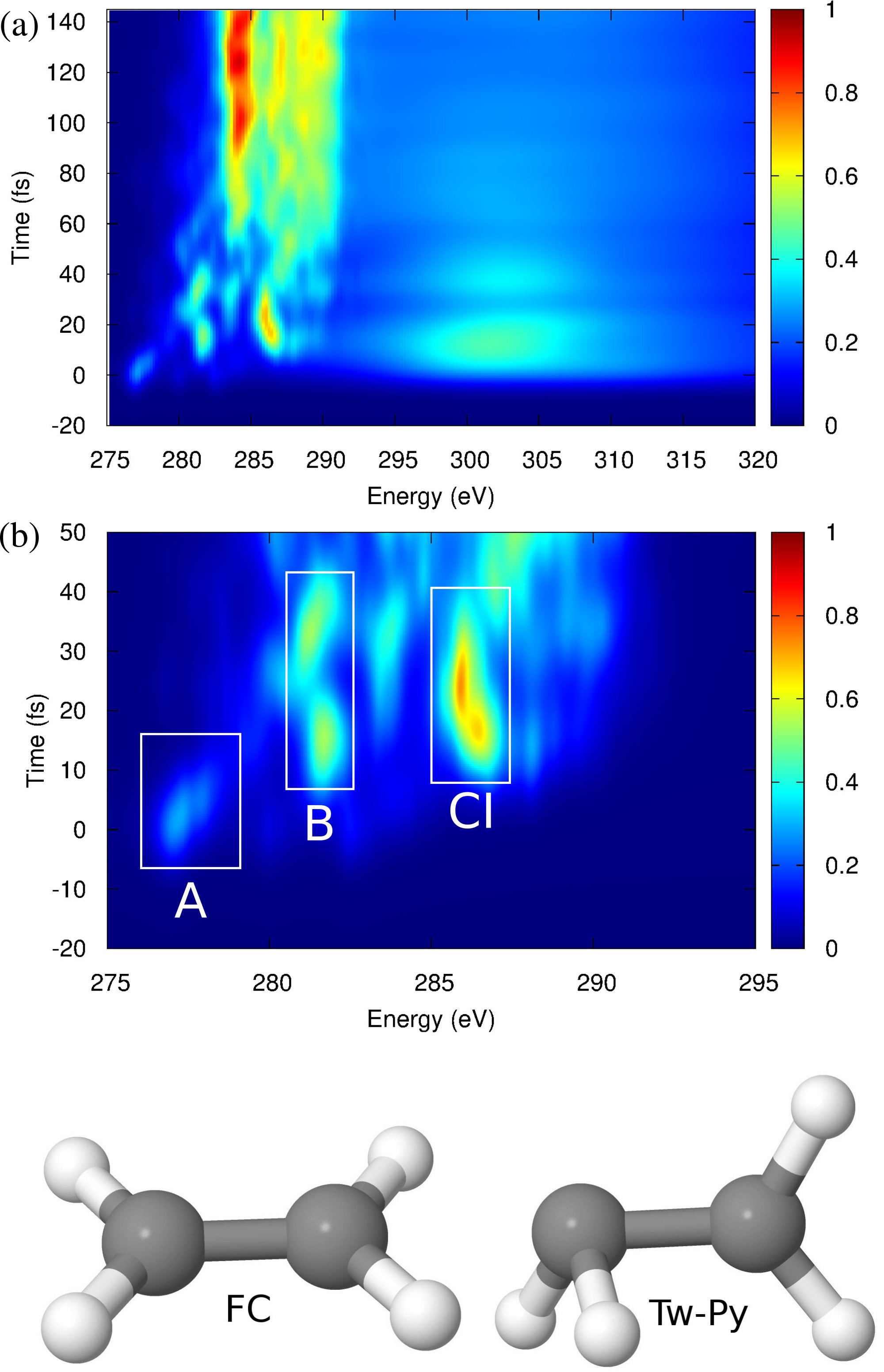}
    \caption{The TRXAS spectrum calculated from AIMS simulations of
      ethylene excited to its $\pi\pi^{*}$ state. (a) The full
      spectrum showing both the pre-edge and post-edge continuum
      absorption. (b) The pre-edge part of the spectrum at short times
      and the relevant peak assignments. The features labeled ``A'',
      ``B'' and ``CI'' are directly related to the associated
      dynamical features shown in Figure \ref{fig:pes}. The FC (``A'')
      and Tw-Py intersection (``CI'') geometries are shown below. The
      color map represents the relative probability of absorption.}
    \label{fig:trxas}
  \end{center}
\end{figure}

In Figure \ref{fig:trxas} (a) we show the time evolution of the C K-edge
XANES spectrum following excitation to the $S_{2}(\pi\pi^{*})$ state,
while Figure \ref{fig:trxas} (b) focuses on the pre-edge region
only ($<$ 295 eV). The post-edge NEXAFS region reflects structural 
dynamics in the excited molecule and will be discussed in a future 
publication. Feature ``A'' at $t$=0 and 277 eV is assigned to excitation of
initially prepared $\pi \pi^{*}$ state to the $1s \pi^{*}$
core-excited state\cite{ethylene_faraday_ours}. The sweep of this
feature to higher energies is a result of rapid twisting about the C-C
bond, lowering the energy of the $\pi \pi^{*}$ state. Feature ``B'',
beginning at around 281 eV and 10 fs, corresponds to two components of
the wavepacket: (i) the portion in the transiently populated $\pi 3s$
state which transitions to the $1s 3s$; (ii) the component
passing through the Tw-Py CI seam. The feature labeled ``CI'', centred
around 286 eV begins to develop intensity at around $\sim$10 fs. It is
also assigned to core-excitation from the $\pi \pi^{*}$
state. Significantly, the ``CI'' feature originates only from those
components of the wavepacket in the $\pi \pi^{*}$ state that are in
close proximity to the Tw-Py CI seam. In other words, the calculated
TRXAS reveals a clear and direct signature of the arrival of the
excited state wavepacket at the CI. Finally, the broad, intense set of
peaks appearing at times $t \gtrsim 50$ fs in Figure \ref{fig:trxas}
(a) corresponds to the vibrationally hot ground electronic state which 
appears following passage through the CI.

The most remarkable feature of the calculated TRXAS is that it clearly encodes the
arrival of the wavepacket at the Tw-Py CI.
Figures S1-S5 in the supplemental material\cite{supp_mat} show the XAS calculated at
two relevant geometries for the states involved and is a convenient
reference for the following discussion. Importantly, we find that at
geometries close to the Tw-Py CI geometry, a remarkable splitting appears 
in the higher energy peak for the $\pi \pi^{*}$ state (corresponding to the $S_{1}$
state at this geometry) pre-edge XAS. At the initial
Franck-Condon (FC), point labeled ``A'', the $\pi \pi^{*}$ state
pre-edge XAS is dominated by a single transition corresponding to the
excitation of a $1s$ core electron into the hole in the
singly-occupied $\pi$ orbital. At the FC point, the molecule possesses
$D_{2h}$ symmetry, meaning that the two $1s$ orbitals are delocalized
across the two indistinguishable C atoms, and only excitation from the
$a_{g}$ $1s$ orbital into the $\pi$ orbital is dipole allowed. That
is, only one $1s \rightarrow \pi$ transition is dipole allowed at this
point. Upon pyramidalization at one of the C atoms, the symmetry is
lowered and the two C atoms now become distinguishable.
The consequences of this are twofold: (i) the $1s$ orbitals now
become localized about the C atoms; (ii) excitation from both of
these $1s$ orbitals into the $\pi$ orbital hole now becomes dipole
allowed. Accordingly, two $1s \rightarrow \pi$ transitions now appear
in the XAS. Importantly, as we discuss below, the splitting between
the two transition energies is quite large, $\sim$4.5 eV, due to the
significant difference in the local valence electronic structure that develops
around the two C atoms. In other words, the charge separation 
driven by the non-adiabatic dynamics is probed in an
atom-specific manner by absorption from each of the now localised, 
distinguishable $1s$ orbitals.

\begin{figure}
  \begin{center}
    \includegraphics[width=7.0cm,angle=0]{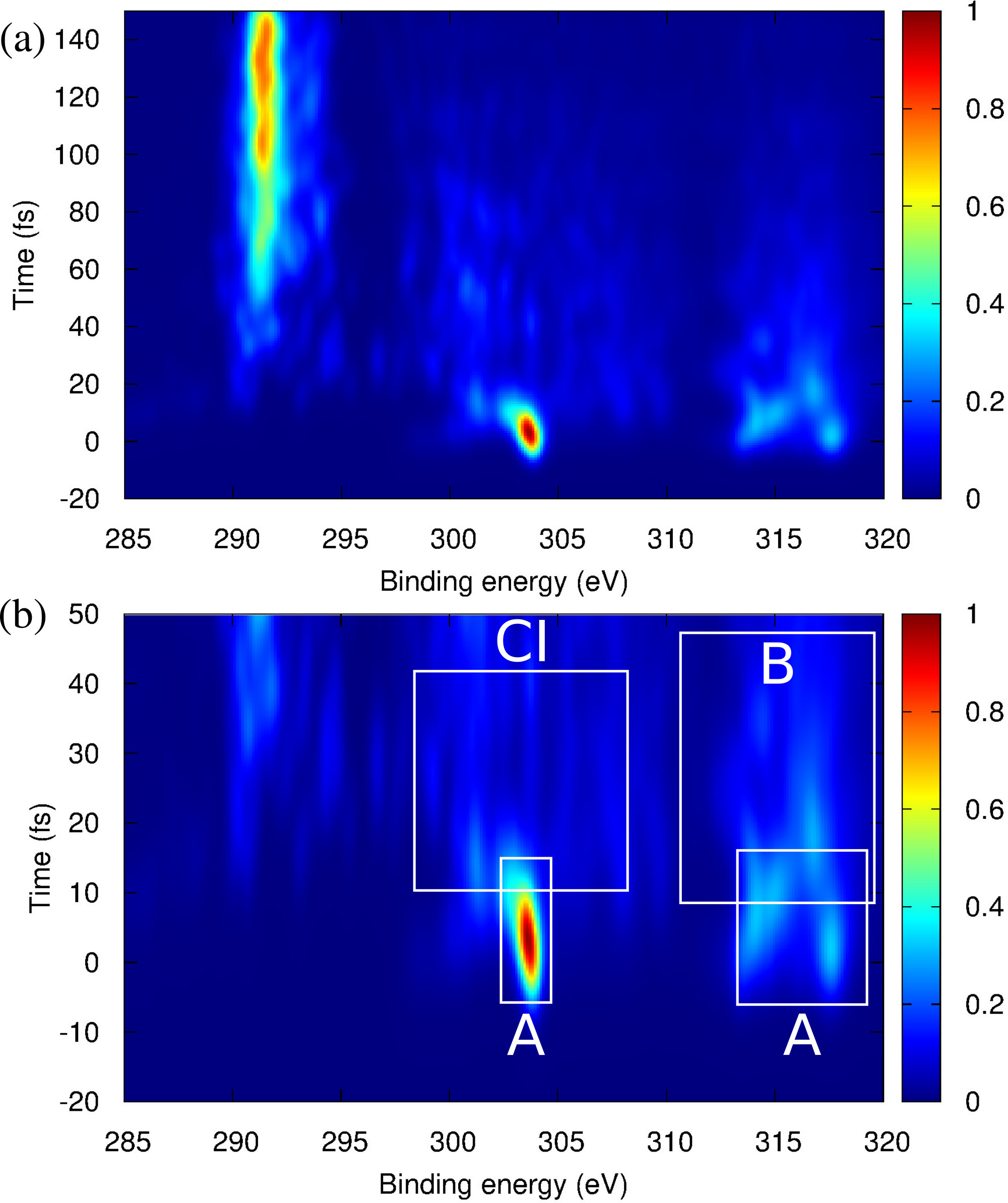}
    \caption{TRXPS calculated using the results of AIMS simulations of
      ethylene excited to the $\pi\pi^{*}$ state. A probe photon
      energy of 320 eV is assumed. (a) The spectrum for all simulation
      times. (b) The spectrum focused on short times. The features
      labeled ``A'', ``B'' and ``CI'' are related to the dynamical
      features shown in Figure \ref{fig:pes}.  The color map
      represents the relative probability of ionization.}
    \label{fig:trxps}
  \end{center}
\end{figure}

Since the nonadiabatic AIMS dynamics simulations which underlie the
TRXAS spectrum are identical to those used in the calculated TRXPS
spectrum, the relative sensitivities of these two techniques to dynamics at
CIs can be compared. In Figure \ref{fig:trxps} we show the AIMS simulation
of the TRXPS spectrum for these same molecular dynamics. At time
$t=0$, the TRXPS is dominated by a single peak corresponding to
ionization of the initially prepared $\pi\pi^{*}$ state in the FC
region. Interestingly, the fingerprints of the dynamical evolution are
somewhat less clear than in the TRXAS.

Figures S6-S10 show the static X-ray photoelectron spectra (320 eV) at
two key geometries (FC ``A'' and Tw-Py ``CI'') for ionization from the
relevant electronic states. At early time delays, feature ``B'' in the
314-318 eV region in Figure \ref{fig:trxps} shows ionization from
$\pi3s$, but is overlapped with allowed ionization channels from the
$\pi\pi^{*}$ FC geometry ``A'', as well ionization from $S_1$ at the
Tw-Py CI. Regarding the latter, the splitting of the two inequivalent
C 1s centres at the Tw-Py CI manifests itself in the splitting between
the group of two peaks centred at 299 eV and the group of two peaks at
306 eV in Figure S9. However, in the wave packet simulation, these
peaks overlap somewhat with the bright peak near 304 eV corresponding
to ionization of the initially prepared $\pi\pi^{*}$, obscuring
somewhat the signature of the CI.

Confirmation of the origin of this large splitting of the core excitation
energies close to the Tw-Py CI comes from consideration of a
one-particle, one-hole (1p1h) excitation operator
$\hat{C}_{\nu c}=\hat{c}_{\nu}^{\dagger} \hat{c}_{c}$ applied to the
correlated ground electronic state $| \Psi_{0} \rangle$ to yield the
singly core-excited configuration
$| \Psi_{\nu c} \rangle = \hat{C}_{\nu c} | \Psi_{0} \rangle$, where
$c$ indexes a core orbital and $\nu$ a virtual valence
orbital. Applying second-order perturbation theory, the
core-excitation energy $\Delta E_{\nu c}$ obtained using the
unperturbed 1p1h configuration $| \Psi_{\nu c} \rangle$ can be written
as\cite{schirmer_adc_orig,schirmer_core_relaxation}:

\begin{equation}\label{eq:2nd-ord_excit_en}
  \Delta E_{\nu c} = \left[ \epsilon_{\nu} - \epsilon_{c} \right] +
  \left[ 2 \langle c \nu | \nu c \rangle - \langle c \nu | c \nu
    \rangle \right] + \Delta E_{\nu c}^{(2)}.
\end{equation}

Here, $\epsilon_{p}$ denotes the energy of the $p$th canonical
Hartree-Fock orbital. At the Tw-Py CI geometry, the splitting of the
two pre-edge $\pi \pi^{*}$ state peaks is accounted for almost
entirely by the zeroth-order term
$\left[ \epsilon_{\nu} - \epsilon_{c} \right]$. Importantly, the
splitting of the two peaks is a consequence of the splitting of the
corresponding $1s$ orbital energies due to the distinguishability of
the two C atoms which occurs uniquely at the CI.

The physical picture emerging from this analysis is that the splitting
of the C 1s peaks is correlated with the onset of charge separation
across the C-C bond (the so-called sudden polarization
effect)\cite{schaefer_twpy_1979} which occurs uniquely at the CI. To
show this, we use the following metric for charge separation,
$\Theta(t)$, across the C-C bond derived from the AIMS simulation:

\begin{equation}
  \Theta(t) = \sum_{I=1}^{n_{s}} \sum_{j=1}^{N_{I}} \left|
    C_{j}^{(I)}(t) \right|^{2} \left| \left\langle g_{j}^{(I)}(t)
      \middle| \left\langle I \middle| \boldsymbol{\mu} \cdot
        \boldsymbol{\nu}_{CC} \middle| I \right\rangle \middle|
      g_{j}^{(I)}(t) \right\rangle \right|.
\end{equation}

\begin{figure}
  \begin{center}
    \includegraphics[width=7.0cm,angle=0]{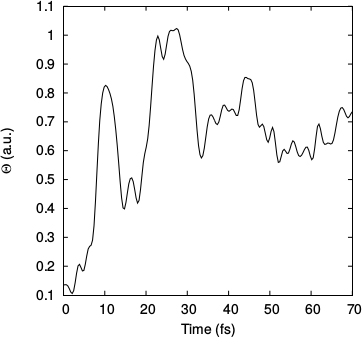}
    \caption{Time-evolution of the charge transfer metric $\Theta$
      calculated using the results of AIMS simulations of ethylene
      excited to the $\pi\pi^{*}$ state. The large, transient charge
      separations occurs when the wavepacket encounters the Tw-Py CI
      and is responsible for the distinguishability of the two C atoms
      at the CI.}
    \label{fig:dipx}
  \end{center}
\end{figure}

\noindent
Here, $\boldsymbol{\mu}$ denotes the molecular dipole operator, and
$\boldsymbol{\nu}_{CC}$ the unit vector coincident with the C-C
bond. The nuclear part of the integral $\left\langle g_{j}^{(I)}
\middle| \left\langle I \middle| \boldsymbol{\mu} \cdot
\boldsymbol{\nu}_{CC} \middle| I \right\rangle \middle| g_{j}^{(I)}
\right\rangle$ is evaluated analytically, whilst the electronic part
is evaluated using a first-order saddle point approximation. The
short-time evolution of $\Theta(t)$ is shown in Figure
\ref{fig:dipx}. It is clear that charge separation across the C-C bond
occurs rapidly, with a local maximum value of $\Theta$ being attained
at around 10 fs. This correlates with the appearance and initial
growth in intensity of the peak centred at 286 eV in the calculated
TRXAS (Figure \ref{fig:trxas}). The onset of charge separation at the
CI leads to the splitting of the peaks in $\pi \pi^{*}$ state pre-edge
XAS. The magnitude of this splitting is a consequence of the
separation of the $1s$ orbital energies, as a result of the transient
charge separation across the C-C bond that occurs uniquely at the CI.
We expect this to be a general feature of dynamics at CIs in
molecules containing C=C double bonds. Such molecules play central 
roles in photochemistry, photobiology and material science.

On the basis of these AIMS simulations, we are emboldened to draw some
conclusions. Firstly, employing core electrons to probe valence
electron density via ultrafast X-ray spectroscopy results in an
exquisitely sensitive measure of complex, dynamic electronic
structures. While both TRXAS and TRXPS encode the nonadiabatic
dynamics, the overlapping continua in the TRXPS may tend to obscure
the CI dynamics. In contrast, TRXAS has fewer dipole-allowed
transitions, leading to a direct mapping of absorption peaks to
specific dynamical features. We conclude that ultrafast X-ray
spectroscopy is a particularly powerful probe of dynamics at conical
intersections.  The commonality of charge separation dynamics
(e.g. proton or electron transfer) in molecular and material processes
suggests that TRXAS and TRXPS will have broad applicability, leading to
a `transition state spectroscopy' for the excited state. Work is
now in progress, within our group and elsewhere, to implement 
these findings experimentally.

Acknowledgements: M.S. and A.S. acknowledge the support of the
National Science and Energy Research Council (Canada) via the
Discovery Grants program. MC acknowledges support of the NCCR:MUST of
the Swiss NSF.\nocite{pickup_photoionisation,spanner_dyson_norms,
roos_anorydberg,columbus,cfour,martinez_ethyleneI,martinez_ethylene_2009}

\bibliography{references}

\end{document}